\documentclass[11pt,a4paper]{scrartcl}

\usepackage{ILD}

\usepackage[symbol]{footmisc}
\usepackage{feynmf}
\usepackage{multirow}
\usepackage{textpos}
\usepackage{lineno}
\usepackage{afterpage}

\RequirePackage{amsthm}
\RequirePackage{newtxmath}

\usepackage{tabularx}
    \newcolumntype{L}{>{\raggedright\arraybackslash}X}

\def\epsilonb{\ensuremath{\epsilon_{b}}\xspace}
\def\epsilonc{\ensuremath{\epsilon_{c}}\xspace}
\def\epsilonuds{\ensuremath{\epsilon_{uds}}\xspace}
\def\epsilonb2{\ensuremath{\epsilon^{2}_{b}}\xspace}
\def\epsilonc2{\ensuremath{\epsilon^{2}_{c}}\xspace}
\def\epsilonuds2{\ensuremath{\epsilon^{2}_{uds}}\xspace}

\def\fb{fb\ensuremath{^{-1}}\xspace}

\def\qqbar{\ensuremath{q}\ensuremath{\overline{q}}\xspace}

\def\ee{\ensuremath{e^{-}e^{+}}\xspace}

\def\cme{\ensuremath{c.m.e.}\xspace}

\def\dEdx{\ensuremath{dE/\/dx}\xspace}
\def\dNdx{\ensuremath{dN/\/dx}\xspace}

\def\AFBb{\ensuremath{A^{b}_{FB}}\xspace}

\def\AFB{\ensuremath{A_{FB}}\xspace}

\def\AFBc{\ensuremath{A^{c}_{FB}}\xspace}

\def\cquark{\ensuremath{c}-quark\xspace}

\def\Zpole{\ensuremath{Z}-pole\xspace}

\def\Pb{\ensuremath{P_{chg.}}\xspace}

\def\Aone{\ensuremath{A_{1}}\xspace}
\def\Atwo{\ensuremath{A_{2}}\xspace}
\def\Athree{\ensuremath{A_{3}}\xspace}

\def\Amodels{\ensuremath{A} models\xspace}

\def\Bmodels{\ensuremath{B} models\xspace}

\def\Bm{\ensuremath{B}\xspace}
\def\BL{\ensuremath{B_{L}}\xspace}
\def\BH{\ensuremath{B_{H}}\xspace}
\def\Bplus{\ensuremath{B_{+}}\xspace}
\def\Bminus{\ensuremath{B_{-}}\xspace}

\def\sinW{\ensuremath{\sin\theta^{0}_{W}}\xspace}
\def\sinW2{\ensuremath{\sin^2\theta^{0}_{W}}\xspace}

\title{Experimental prospects for indirect BSM searches in  $e^{-}e^{+}\rightarrow q\bar{q}$ ($q=c,b$)  processes at Higgs Factories}
\titlecomment{This work was carried out in the framework of the ILD concept group}
\titlecomment{The European Physical Society Conference on High Energy Physics (EPS-HEP2023), 21-25 August 2023, Hamburg, Germany}

\ildproc{PHYS}{2023}{013, IFIC/23-51} 

\date{\today}

\addauthor{J.P. M\'arquez}{\institute{1} 
\footnote{Speaker \href{mailto:jesus.marquez@ific.uv.es}{jesus.marquez@ific.uv.es}}}

\addinstitute{1}{IFIC, Universitat de Val\`encia and CSIC, C./ Catedr\'atico Jos\'e Beltr\'an 2, E-46980 Paterna, Spain}

\abstract{
    This contribution explores the ability to probe BSM physics by using the experimental prospects for measuring the forward-backward asymmetry ($A_{FB}$) in $e^{+}e^{-}\rightarrow b\bar{b}$ and $e^{+}e^{-}\rightarrow c\bar{c}$ processes at the baseline energy points of ILC: 250 and 500 GeV. The studies are based on the full simulation samples and reconstruction chains from the ILD concept group. The BSM models studied are two different types of gauge-Higgs unification (GHU) models that predict BSM Z$^\prime$ resonances at the TeV scale.
}

\graphicspath{ {./logos/}{./figures/} }

\begin{document}

\titlepage


\section{Introduction}

In the Standard Model (SM) of particle physics, the Higgs boson does not follow the gauge principle as the rest of the SM bosons do. As a result, the values of the Higgs couplings with quarks and leptons, as well as its self-coupling, are fixed parameters. Additionally, the Higgs mass adquires large quantum corrections that have to be canceled by \textit{fine-tuning}. A possible solution for the aforementioned inconsistencies is achieving stabilization of the Higgs mass against quantum corrections by extending the gauge group of the SM. These models are referred to as Gauge-Higgs Unification (GHU) models. 

In this document, a first approach to the phenomenology of GHU models will be addressed in the context of a future \textit{Higgs factory}: The International Linear Collider (ILC)~\cite{Behnke:2013xla,Baer:2013cma,Adolphsen:2013jya,Adolphsen:2013kya,Behnke:2013lya}. This study has been done featuring the International Large Detector (ILD)~\cite{Behnke:2013lya,ILD:2020qve} at ILC with full simulation studies at 250 and 500 GeV collision energies and using the forward-backward asymmetry (\AFB) in $e^{+}e^{-}\rightarrow b\bar{b}$ and $e^{+}e^{-}\rightarrow c\bar{c}$ processes as the observable to discriminate these models. The \AFB is defined as:
\begin{equation}
A^{q\bar{q}}_{FB}=\frac{\sigma^{F}_{e{-}e^{+}\rightarrow q\bar{q}}-\sigma^{B}_{e{-}e^{+}\rightarrow q\bar{q}}}{\sigma^{F}_{e{-}e^{+}\rightarrow q\bar{q}}+\sigma^{B}_{e{-}e^{+}\rightarrow q\bar{q}}}
\label{formula:AFB}
\end{equation}
where $\sigma^{F/B}_{e{-}e^{+}\rightarrow q\bar{q}}$ is the cross-section in the forward/backward hemisphere as defined by the polar angle $\theta_q$.

\section{Gauge-Higgs Unification (GHU) models}
\label{Sec:GHU}
Gauge-Higgs unification (GHU) introduces the Higgs boson as an Aharonov-Bohm (AB) phase $\theta_{H}$ in an extra dimension. The GHU models extend the gauge theory of the SM to $SU(3)_{c} \times SO(5)_W \times U (1)_{x}$ in a Randall-Sundrum warped space\cite{Funatsu:2017nfm,Funatsu:2020haj}. In this context, gauge symmetry stabilizes the Higgs-boson mass against quantum corrections. At low energies the phenomenology of GHU matches the SM but the prediction of existence of $Z^{\prime}$ bosons produces deviations at high energies. The $Z^{\prime}$ bosons predicted by GHU are Kaluza-Klein (KK) resonances of $\gamma$, $Z$, and $Z_R$. The angle $\theta_{H}$ and KK mass $m_{KK}$ and are set in a way that reproduces the masses and decay width of the SM $W$ and $Z$ bosons, which are the zeroth mode of the 5 dimensional fields. In this work, two different types of GHU models are considered: \Amodels \cite{Funatsu:2017nfm} and \Bmodels \cite{Funatsu:2020haj}. 

In ref.~\cite{Funatsu:2017nfm}, three variations of the \Amodels (\Aone, \Atwo, \Athree) are proposed (see Tab.~\ref{table:A}).
In these models, quark-lepton multiplets are introduced in the vector representation of $SO(5)$. 
These models use $\theta_{H}\simeq 0.11-0.07$ and KK resonances masses $m_{KK}\simeq 8-11$ TeV. The non-observation in LHC for direct measurements of $Z^{\prime}$ set a limit of $\theta_{H}\lesssim 0.11$. 
They also predict large couplings of the right-handed (down-type) fermions to the $Z^{\prime}$ resonances. 
In all \Amodels, the parameters are chosen such that $Z$ couplings to fermions (except top-quark) agree with the SM within one part in $10^4$. 
\begin{table}[!ht]
  \centering
\begin{tabular}{|c|c|c|c|c|c|c|c|c|}
\hline
Model & \begin{tabular}[c]{@{}c@{}}$\theta_H$\\ {[}rad.{]}\end{tabular}  & \begin{tabular}[c]{@{}c@{}}$m_{KK}$\\ {[}TeV{]}\end{tabular} & \begin{tabular}[c]{@{}c@{}}$m_{Z^{(1)}}$\\ {[}TeV{]}\end{tabular} & \begin{tabular}[c]{@{}c@{}}$\Gamma_{Z^{(1)}}$\\ {[}TeV{]}\end{tabular} & \begin{tabular}[c]{@{}c@{}}$m_{\gamma^{(1)}}$\\ {[}TeV{]}\end{tabular} & \begin{tabular}[c]{@{}c@{}}$\Gamma_{\gamma^{(1)}}$\\ {[}TeV{]}\end{tabular} & \begin{tabular}[c]{@{}c@{}}$m_{Z_R^{(1)}}$\\ {[}TeV{]}\end{tabular} & \begin{tabular}[c]{@{}c@{}}$\Gamma_{Z_R^{(1)}}$\\ {[}TeV{]}\end{tabular} \\ \hline
\Aone & 0.115 & 7.41 & 6.00 & 0.406 & 6.01 & 0.909 & 5.67 & 0.729 \\ \hline
\Atwo & 0.0917  & 8.81 & 7.19 & 0.467 & 7.20 & 0.992 & 6.74 & 0.853 \\ \hline
\Athree & 0.0737 & 10.3 & 8.52 & 0.564 & 8.52 & 1.068 & 7.92 & 1.058  \\ \hline
\end{tabular}
  \caption{Parameter setting, masses and decay widths of $Z^{\prime}$ bosons ($Z^{(1)}$, $\gamma^{(1)}$ and $Z_R^{(1)}$) in the \Amodels\label{table:A}}
\end{table}

In ref.~\cite{Funatsu:2020haj}, five variations of the \Bmodels (\Bm, \BL, \BH, \Bminus, \Bplus) are proposed (see Tab.~\ref{table:B}). 
The \Bmodels are inspired by Grand Unification Theories, leading to quark-lepton multiples introduced in the spinor, vector and singlet representations of $SO(5)$. 
The parameters are fixed in a way that predicts values of $A_{FB}(\ee\rightarrow\mu^{-}\mu^{+})$ compatible with current measurements. 
For all \Bmodels, the parameters are chosen such that $Z$ couplings to fermions (except top-quark) agree with the SM within one part in $10^3$. 
The \Bmodels, in contrast with the \Amodels, predict a large deviation for the couplings of the left-handed (up-type) fermions to the new bosons.

\begin{table}[!ht]
  \centering
\begin{tabular}{|c|c|c|c|c|c|c|c|c|c|}
\hline
Model & \begin{tabular}[c]{@{}c@{}}$\theta_H$\\ {[}rad.{]}\end{tabular} & \begin{tabular}[c]{@{}c@{}}$m_{KK}$\\ {[}TeV{]}\end{tabular} & \begin{tabular}[c]{@{}c@{}}$m_{Z^{(1)}}$\\ {[}TeV{]}\end{tabular} & \begin{tabular}[c]{@{}c@{}}$\Gamma_{Z^{(1)}}$\\ {[}TeV{]}\end{tabular} & \begin{tabular}[c]{@{}c@{}}$m_{\gamma^{(1)}}$\\ {[}TeV{]}\end{tabular} & \begin{tabular}[c]{@{}c@{}}$\Gamma_{\gamma^{(1)}}$\\ {[}TeV{]}\end{tabular} & \begin{tabular}[c]{@{}c@{}}$m_{Z_R^{(1)}}$\\ {[}TeV{]}\end{tabular} & \begin{tabular}[c]{@{}c@{}}$\Gamma_{Z_R^{(1)}}$\\ {[}TeV{]}\end{tabular} \\ \hline
\Bm & 0.10 & 13.00  & 10.20 & 8.713 & 10.20 & 3.252 & 9.951 & 0.816 \\ \hline
\BL & 0.10  & 11.00 & 8.713 & 4.773 & 8.715 & 2.080 & 8.420 & 0.603 \\ \hline
\BH & 0.10  & 15.00 & 11.69 & 11.82 & 11.69 & 4.885 & 11.48 & 1.253 \\ \hline
\Bminus & 0.09 & 13.00 & 10.26 & 6.413 & 10.26 & 2.723 & 9.951 & 0.732 \\ \hline
\Bplus & 0.11  & 13.00 & 10.15 & 9.374 & 10.15 & 3.836 & 9.951 & 0.924 \\ \hline
\end{tabular}
  \caption{Parameter setting, masses and decay widths of $Z^{\prime}$ bosons ($Z^{(1)}$, $\gamma^{(1)}$ and $Z_R^{(1)}$) in the \Bmodels\label{table:B}}
\end{table}

\section{Experimental prospects for \AFB at ILC250 and ILC500}
\label{Sec:Exp}
The ILC run plan considered in this study is the Horizon-2020 (H20) as detailed in \cite{Bambade:2019fyw} which foresees $e^{+}e^{-}$ collisions at different center of mass energies ranging from the Z-pole (Giga-Z) up to 1 TeV. The nominal program has an initial stage at 250 GeV (ILC250), including a luminosity upgrade, and a following stage at 500 GeV (ILC500), after an energy upgrade. ILC also features longitudinal polarization for both the electron and positron beams. For H20, the ILC250 physics program foresees a total integrated luminosity of 2000 \fb distributed between four different beam polarization schemes: $45\%$ in $P_{\mathrm{e^{-}e^{+}}}=(-0.8,+0.3)$, $45\%$ in $P_{\mathrm{e^{-}e^{+}}}=(+0.8,+0.3)$, $5\%$ in $P_{\mathrm{e^{-}e^{+}}}=(-0.8,-0.3)$ and $5\%$ in $P_{\mathrm{e^{-}e^{+}}}=(+0.8,-0.3)$. The ILC500 program foresees a total integrated luminosity of 4000 \fb distributed among the different polarization as $40\%$, $40\%$, $10\%$ and $10\%$, with the same polarization configurations as the ILC250 case.

The ILD is one of the two proposed detector concepts for the ILC. It is a highly hermetic multi-purpose detector designed for the maximal exploitation of particle flow techniques in event reconstruction. 
The ILD layout, from the IP to the outside consists of: a high-precision vertex detector (VTX), silicon tracking systems, a time projection chamber (TPC), a highly granular calorimeter system (ECAL and HCAL) and a muon catcher. All the aforementioned subdetectors are placed inside a solenoid providing a magnetic field of $3.5$\,T, surrounded by an iron yoke instrumented for muon detection. The ILD TPC \cite{thelctpccollaboration2016time,LCTPC:2022pvp} allows continuous 3D tracking and particle identification (PID).

The full simulations for this study have been performed via \texttt{ILCSOFT} \textit{v02-02-03}\footnote{Link: \url{https://github.com/iLCSoft}}, which merge different software packages and algorithms that operate in a modular way: from MC events to final reconstruction. All simulations use the ILD-L\cite{ILD:2020qve} model, whose geometry, material and readout systems are implemented in the \texttt{DD4HEP} framework\cite{Frank:2014zya}, interfaced with \texttt{Geant4} toolkit. Both signal and background events from QED ISR are generated with the \texttt{WHIZARD}\cite{Kilian:2007gr} event generator at leading order. The beam energy spectrum and beam-beam interaction is generated via \texttt{Guinea-Pig}\cite{Schulte:1999tx}. The non-perturbative effects of hadronization are provided by the \texttt{Pythia} event generator\cite{Sj_strand_2006}, as well as the final state QCD and QED radiations.
The signal and background cross sections are listed in the references: Tables 1 and 2 from \cite{Irles:2023ojs} for ILC250 and in Tables 1 and 2 from \cite{Irles:2023nee} for ILC500. 
Only backgrounds leading to fully hadronic final states are considered. 
Backgrounds involving leptons in the final states are ignored since those are expected to be easily identified. 
The simulations feature fully polarized beams. 

The reconstruction of the events begins with the track reconstruction performed by the \texttt{MarlinTrk} framework. 
Later, \texttt{Pandora}\cite{Marshall:2015rfa} runs the particle flow algorithm (PFA) that matches the tracking with the high-granular calorimetry information building the particle flow objects (PFO), which are then treated as single particles.
By using \texttt{LCFI+} software tool the PFOs are merged into vertices and jets. Then, also in LCFI+, the flavor tagging of the jets is performed.


All the methods (Preselection, Double tag, Double charge and fitting) for reconstructing the \qqbar system have been developed for the analysis at 250 GeV and are explained in detail in \cite{Irles:2023ojs} together with a comprehensive study of the most dominant systematic uncertainties.
The same methods are used for the study at 500 GeV shown in \cite{Irles:2023nee}. Additionally, in this later document, the improvement gain by using cluster counting techniques (\dNdx) instead of the traditional energy loss (\dEdx) approach was reviewed.


\section{Discrimination power for GHU at ILC250/500}
\label{Sec:DisPow}
Through the detailed studies described in the previous section, we obtain a realistic estimation of the uncertainties on \AFB for the $b$ and \cquark cases at ILC250 and ILC500, with existing detector models and reconstruction tools.
Systematic uncertainties are being ignored for the analysis discussed in this section since \AFB measurements above the \Zpole are expected to be statistically dominated at ILC, as shown in \cite{Irles:2023ojs} and \cite{Irles:2023nee}.
The uncertainties are considered as normally distributed and with no correlations between measurements. 
This second approximation is motivated by the nature of the analysis, given that the Double Tag and Double Charge methods lead to a selection of fully independent samples for the different flavors and polarization. Also, statistically independent MC simulations have been used to analyze the various polarization scenarios.

The statistical significance ($\sigma$-level) when comparing two models, $i$ and $j$, is defined as
\begin{equation}
    d_{ij}=\frac{|A_{FB,i}-A_{FB,j}|}{\Delta A_{FB,j}}
\end{equation}
with $A_{FB,i/j}$ being the \AFB predicted at leading order by the model $i$ or $j$, introduced in Sec.~\ref{Sec:GHU}. 
The $\Delta A_{FB,i/j}$ corresponds to the expected 
statistical uncertainty of the forward-backward measurement at ILC, obtained as explained in Sec.~\ref{Sec:Exp}.
\begin{figure}[!ht]
\begin{center}
    \begin{tabular}{cc}
      \includegraphics[width=0.45\textwidth]{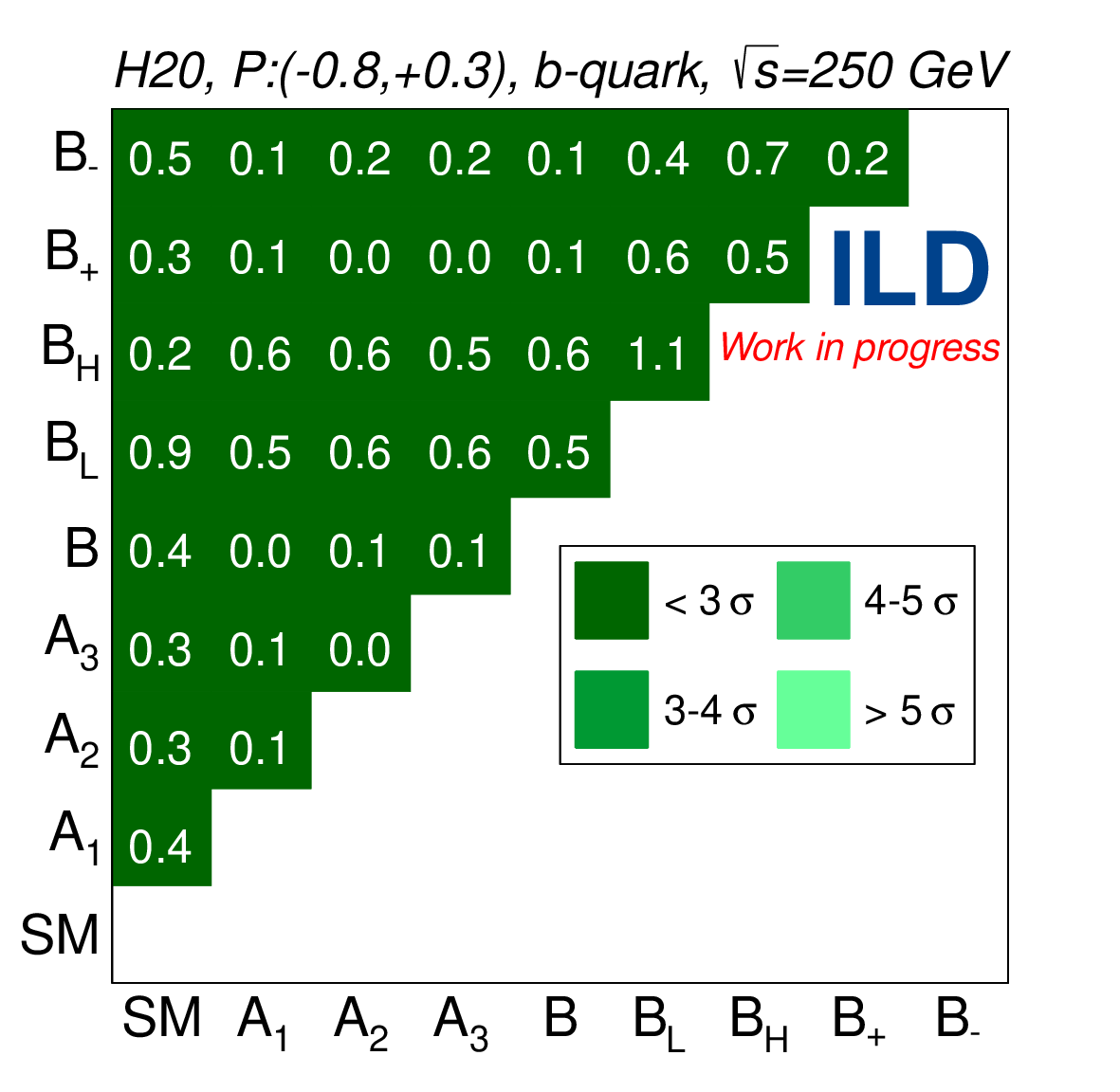} &
      \includegraphics[width=0.45\textwidth]{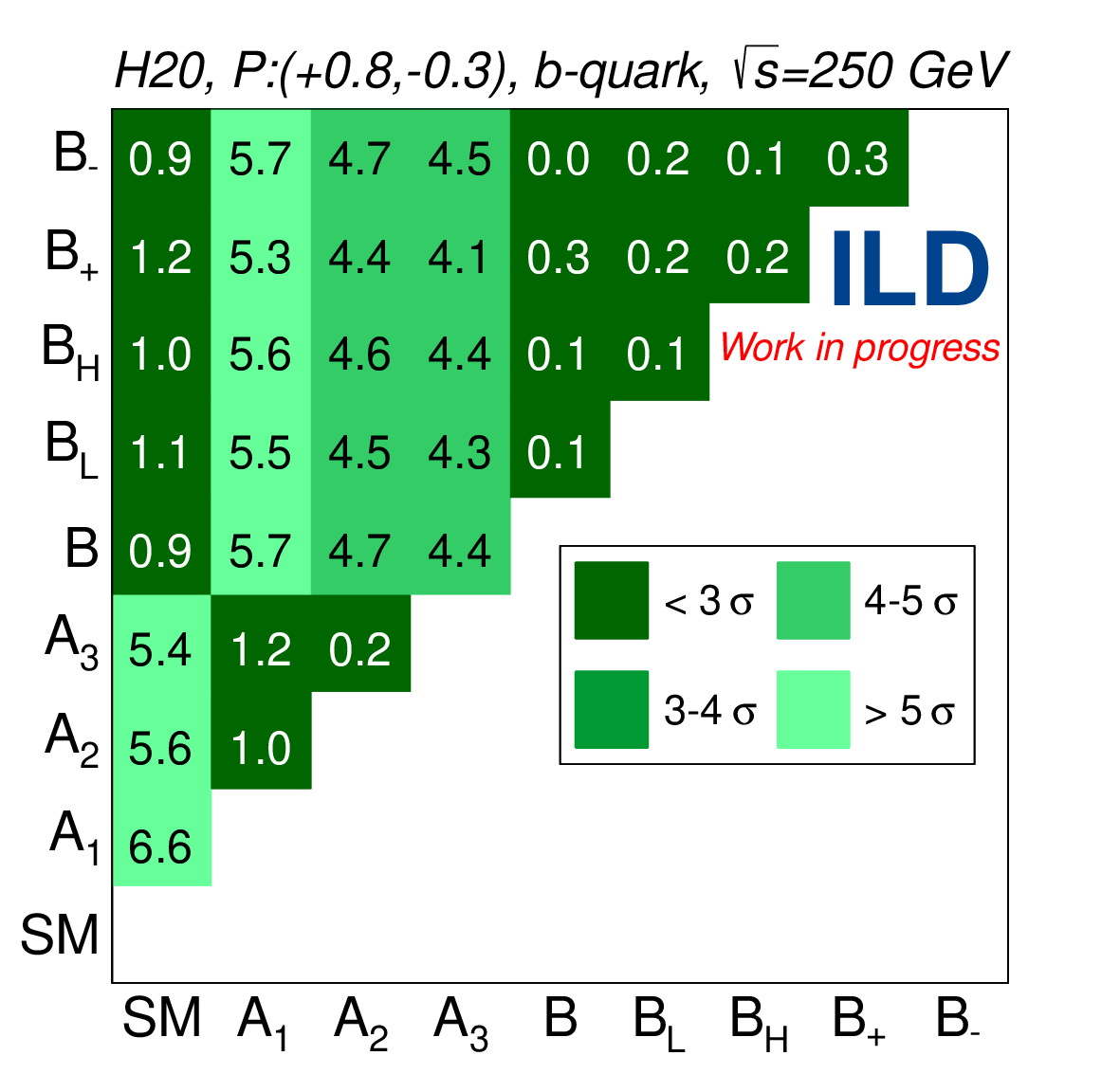} \\
      \includegraphics[width=0.45\textwidth]{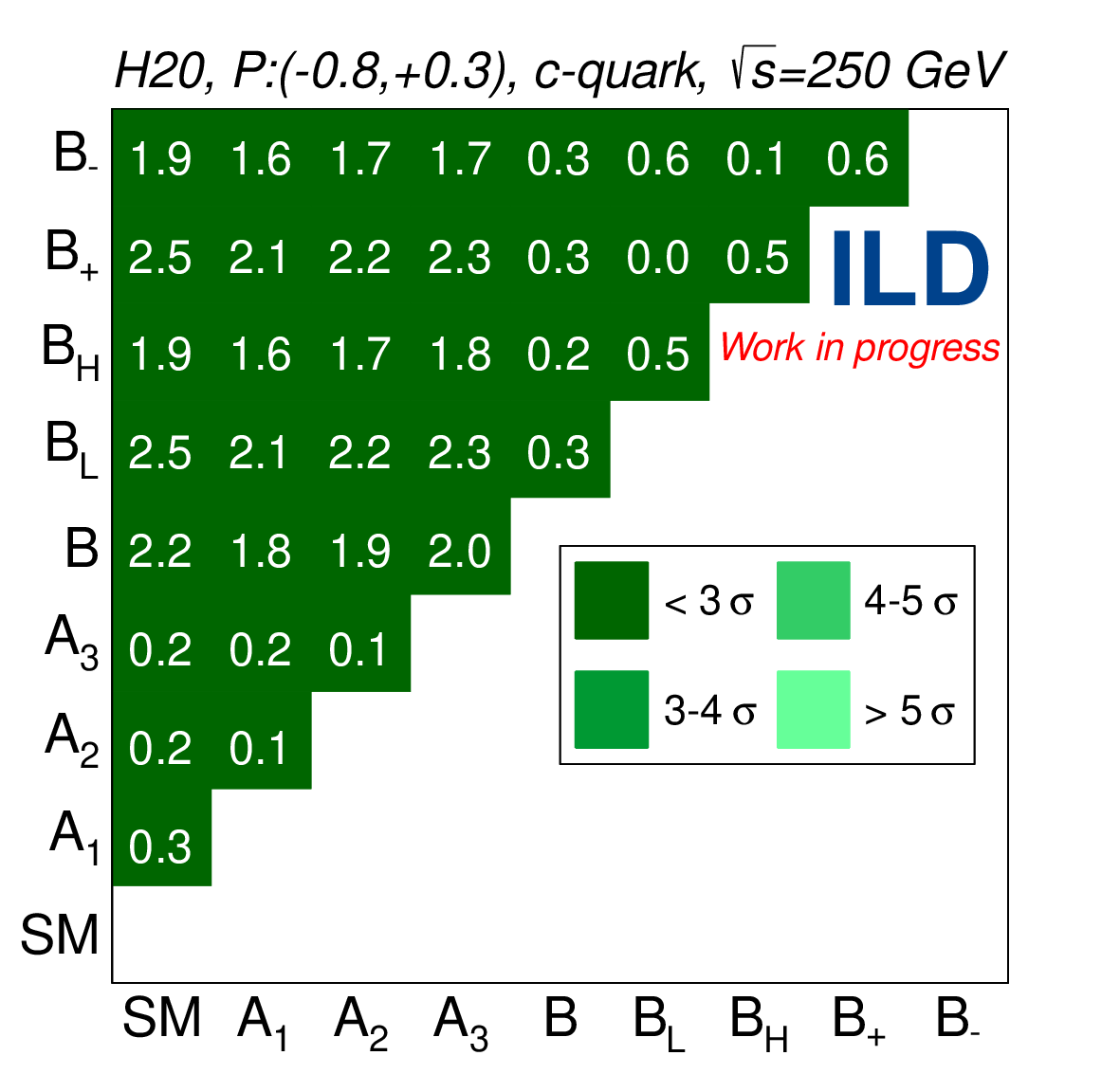} &
      \includegraphics[width=0.45\textwidth]{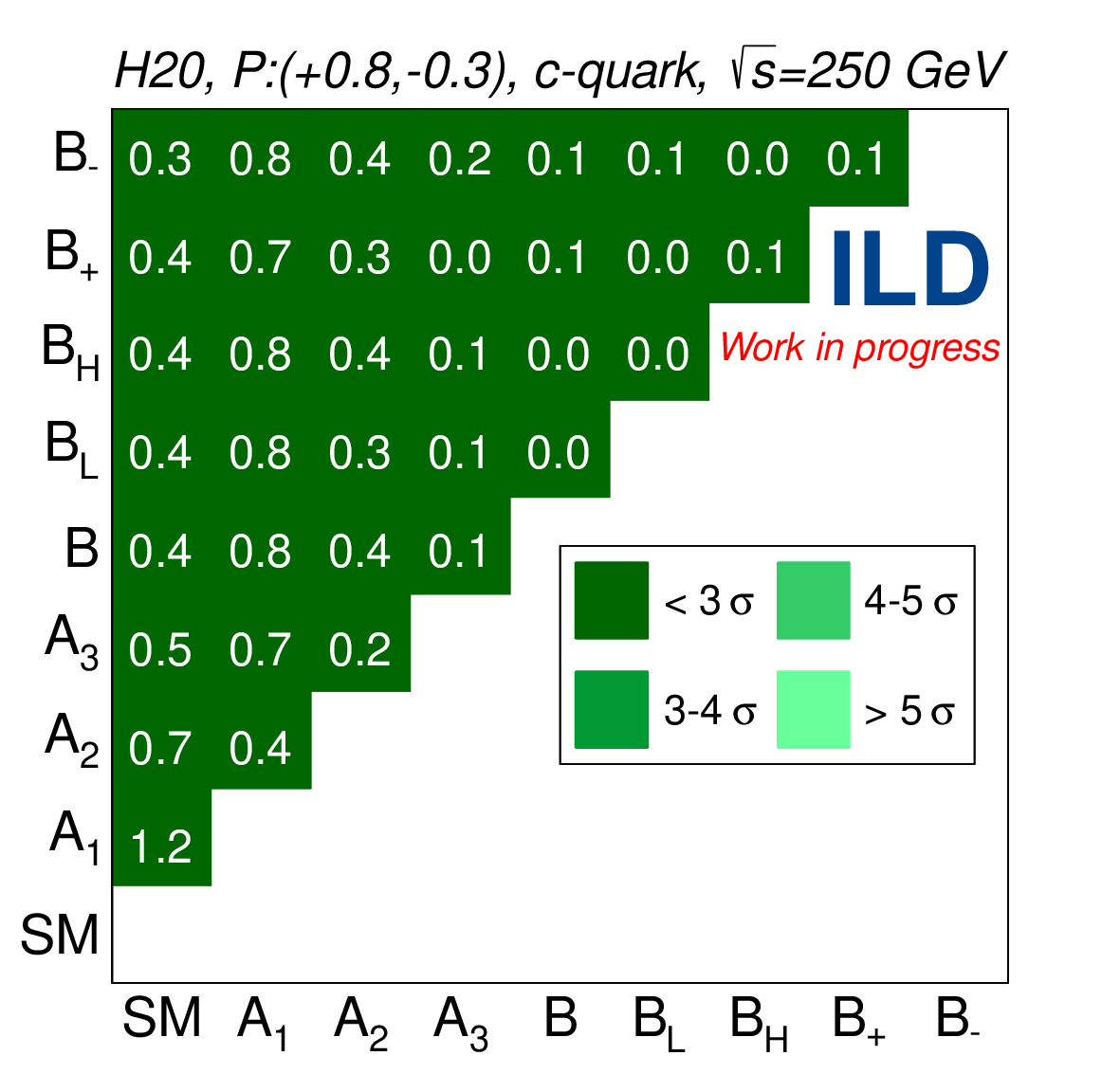} 
    \end{tabular}
\caption{Statistical discrimination power between SM and GHU models using \AFB at ILC250 under the H20 scenario. Plots assuming only statistical uncertainties and the use of \dNdx for PID.\label{Fig:AFB_250_dNdx}}
\end{center}
\end{figure}
\begin{figure}[!ht]
\begin{center}
    \begin{tabular}{cc}
      \includegraphics[width=0.45\textwidth]{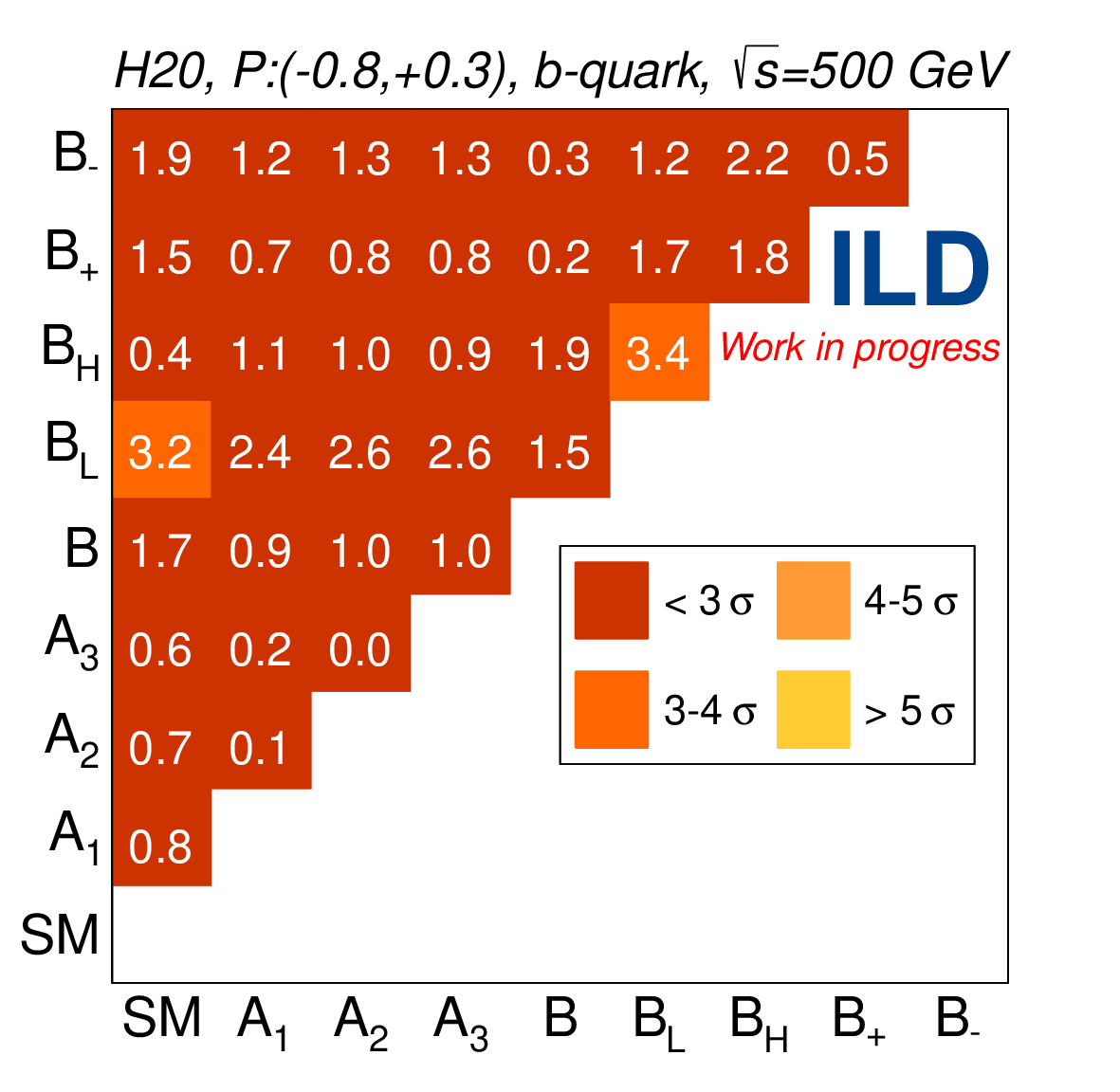} &
      \includegraphics[width=0.45\textwidth]{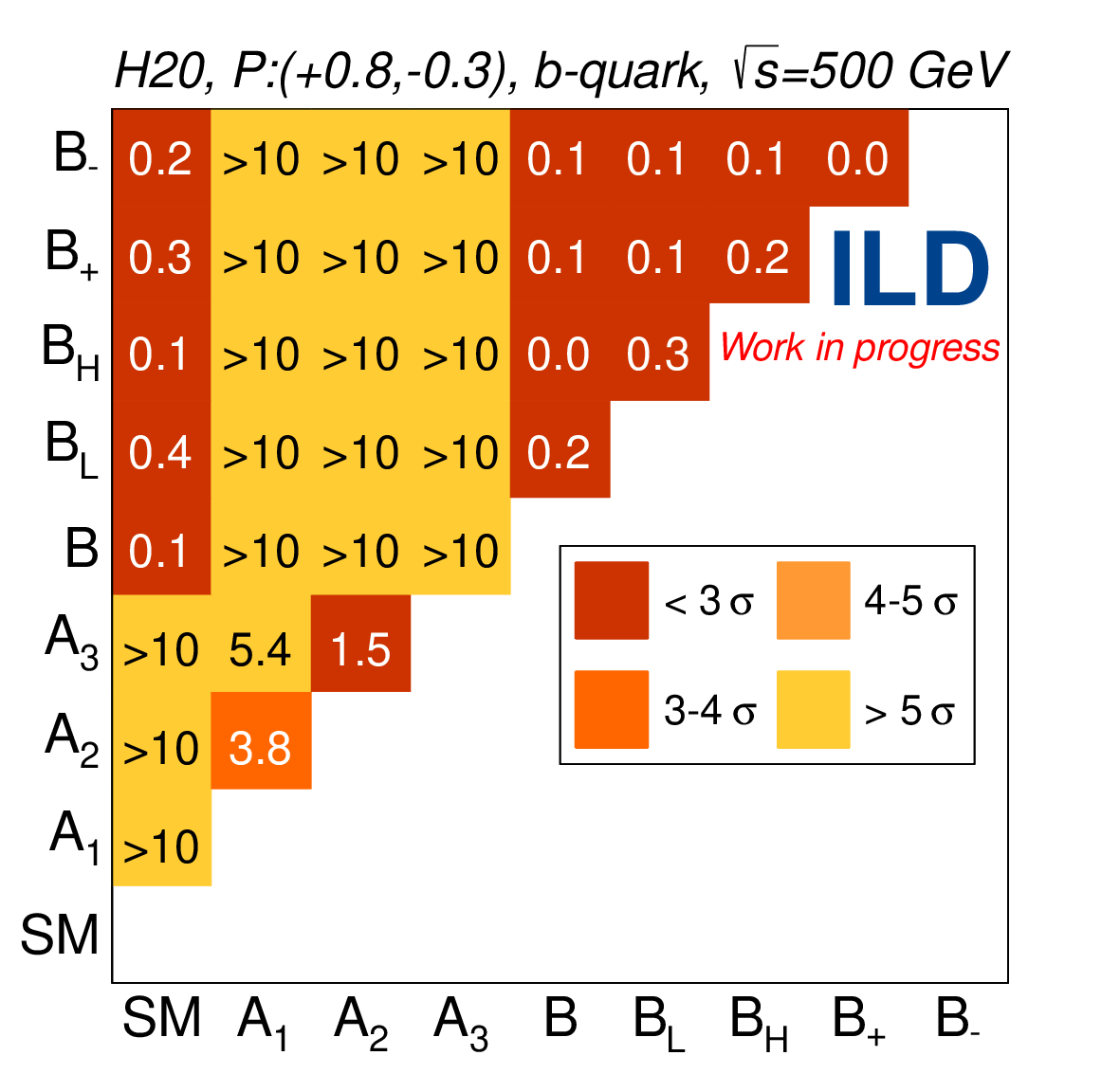} \\
      \includegraphics[width=0.45\textwidth]{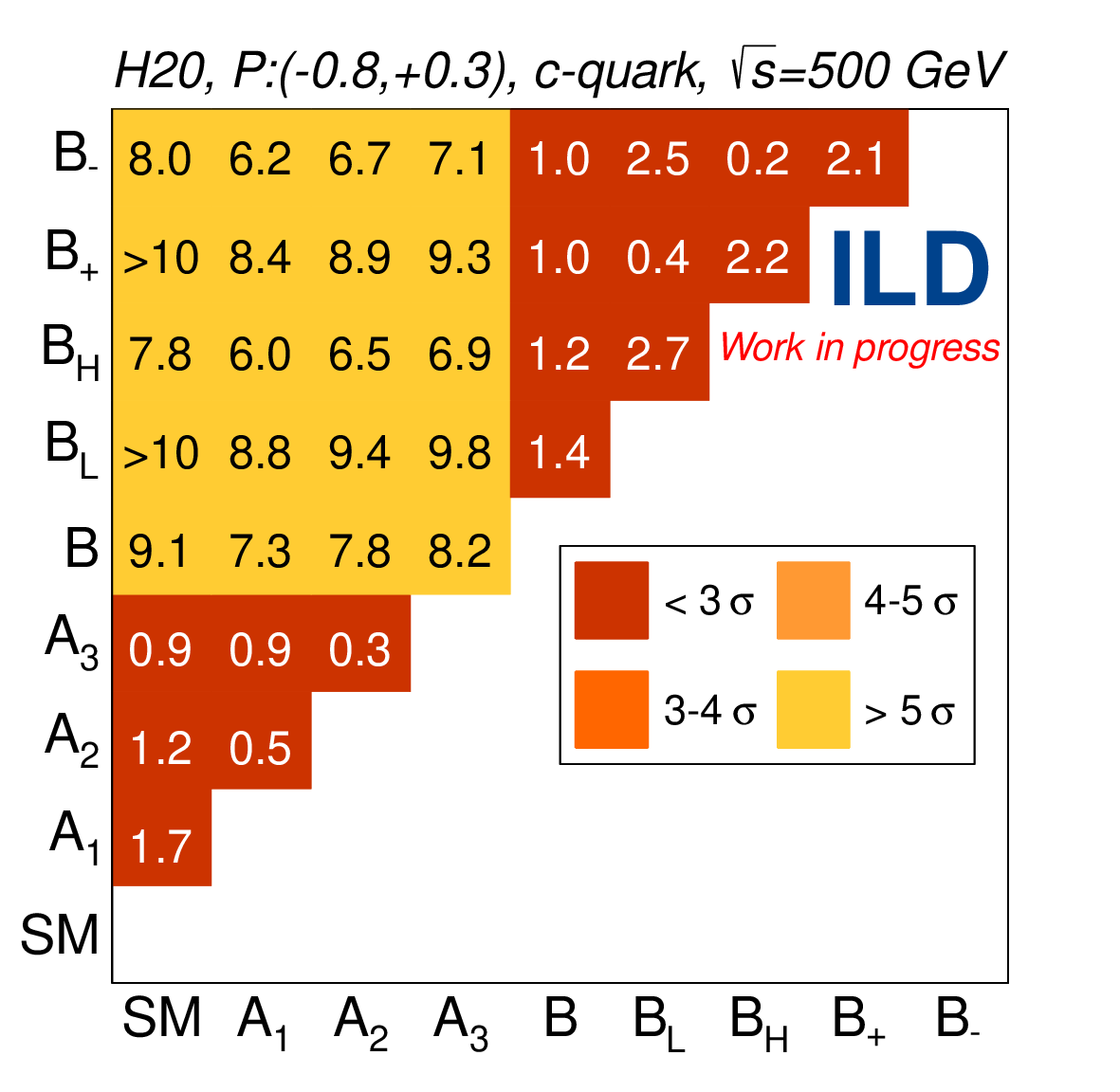} &
      \includegraphics[width=0.45\textwidth]{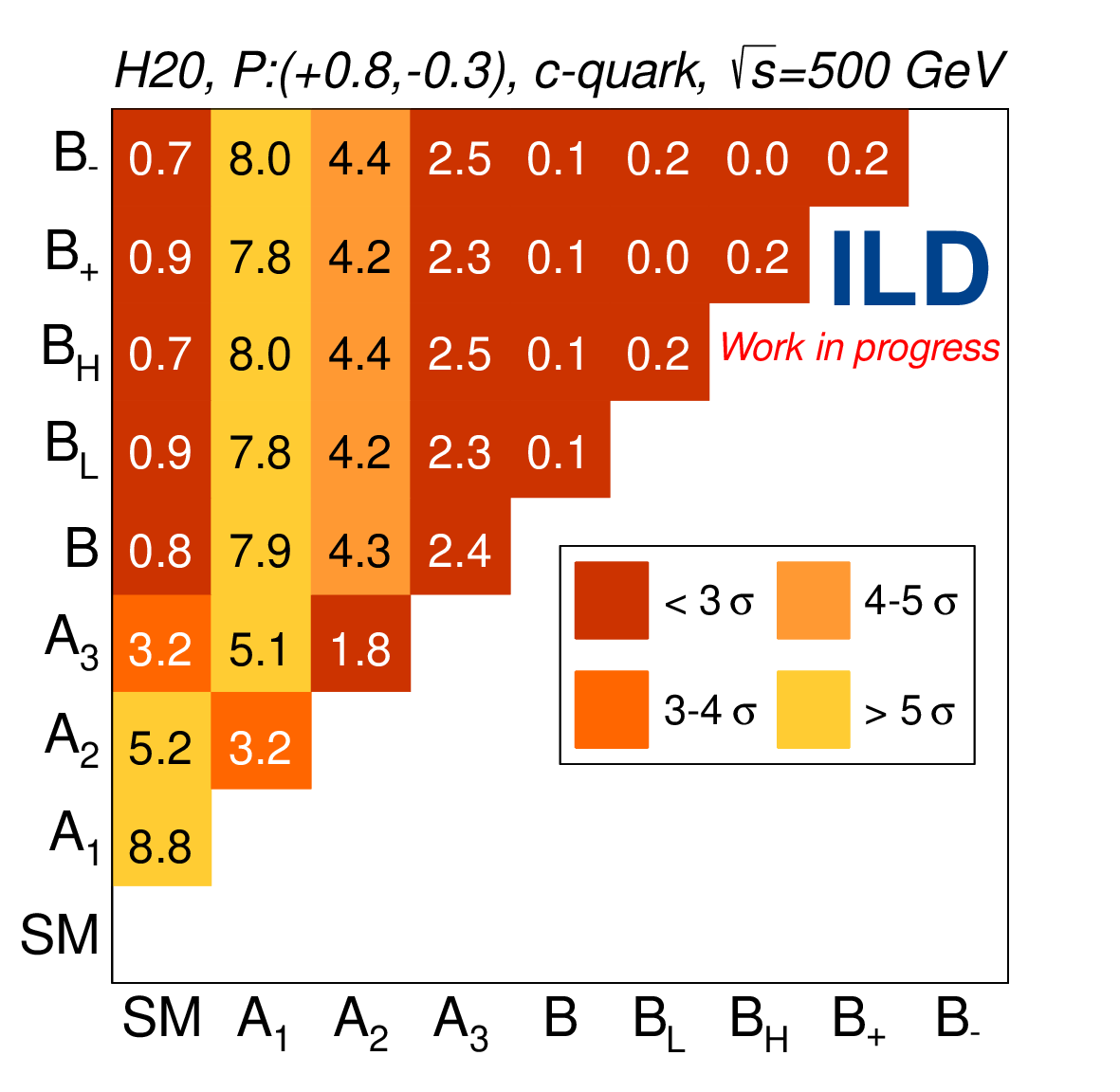} 
    \end{tabular}
\caption{Statistical discrimination power between SM and GHU models using \AFB at ILC500 under the H20 scenario. Plots assuming only statistical uncertainties and the use of \dNdx for PID.\label{Fig:AFB_500_dNdx}}
\end{center}
\end{figure}
In Fig.~\ref{Fig:AFB_250_dNdx} the four \AFB observables at 250 GeV \cme are shown. In this plot is clear how the \AFBb with right-handed electrons and left-handed positrons ($P_{\mathrm{e^{-}e^{+}}}=(+0.8,-0.3)$) collisions provide the best resolution for the \Amodels, so that they could be discriminate from the SM by only this measurement. Given the expected discrimination power, in case of a measurement compatible with any of the \Amodels all of the \Bmodels could be ruled out.

Alternatively, in Fig.~\ref{Fig:AFB_500_dNdx} the \AFB observables at ILC500 are plotted. This plot made manifest the aforementioned sensibility of the \Amodels to the $P_{\mathrm{e^{-}e^{+}}}=(+0.8,-0.3)$ case, and now also for \AFBc. Furthermore, at this energy, the \Bmodels could also be distinguished from the SM by its sensibility to \AFBc for $P_{\mathrm{e^{-}e^{+}}}=(-0.8,+0.3)$.
\section{Conclusions}
This document shows the first insights into the potential of ILC for measuring GHU models by using heavy quark production in its two main stages: ILC250 and ILC500. All the prospects have been done assuming only statistical uncertainties. The \Amodels could be ruled out by measuring \AFBb with $P_{\mathrm{e^{-}e^{+}}}=(+0.8,-0.3)$ at ILC250 whereas accessing to higher energies at ILC500 in mandatory to rule out the \Bmodels, that could be performed by using the \AFBc in $P_{\mathrm{e^{-}e^{+}}}=(-0.8,+0.3)$ beam operation. However, full between-model separation is not possible under these assumptions only using these channels.

The statistical combination of these measurements to obtain a greater discrimination power as well as using other running scenarios such as a 1 TeV ILC run is work yet to be explored. Additionally, other models predicting heavy Z$^{\prime}$ resonances could also be studied in the future, including other GHU models~\cite{Funatsu:2019xwr} or different ones such as the minimal $U(1)_X$ models~\cite{Das_2022}.

\section{Acknowledgments}
We would like to thank the LCC generator working group and the ILD software working group for providing the simulation and reconstruction tools and producing the Monte Carlo samples used in this study.
This work has benefited from computing services provided by the ILC Virtual Organization, supported by the national resource providers of the EGI Federation and the Open Science GRID.
This work was supported by the PlanGenT program from the Generalitat Valenciana (Spain) with the grant number CIDEGENT/2020/21 and by CSIC via the mobility grant iMOVE23245.

\newpage%
\clearpage

\printbibliography[title=References]
\end{document}